# Anisotropic ultrafast spin/valley dynamics in $WTe_2$ films

Yequan Chen[1], Zhendong Chen[1], Yongda Chen[1], Liming Chen[1], Jiai Ning[1], Bo Liu[1], Chunchen Zhang[2], Xianyang Lu[1,3], Xuezhong Ruan[1], Wenqing Liu[4], Peng Wang[2], Fengqi Song[5], Chunfeng Zhang[5], Fengqiu Wang[1], Jing Wu[3], Liang He[1], Xuefeng Wang[1,*], Rong Zhang[1] and Yongbing Xu[1,3,#]

[1]*National Laboratory of Solid State Microstructures and Jiangsu Provincial Key Laboratory of Advanced Photonic and Electronic Materials, School of Electronic Science and Engineering, Nanjing University, Nanjing 210093, China*

[2]*College of Engineering and Applied Sciences, Nanjing University, Nanjing 210093, China*

[3]*York-Nanjing Joint Center in Spintronics, Department of Electronic Engineering and Department of Physics, The University of York, York YO105DD, United Kingdom*

[4]*Department of Electronic Engineering, Royal Holloway University of London, Egham, Surrey TW200EX, United Kingdom*

[5]*School of Physics, Nanjing University, Nanjing 210093, China*

$WTe_2$ Weyl semimetal hosts the natural broken inversion symmetry and strong spin orbit coupling, making it promising for exotic spin/valley dynamics within a picosecond timescale. Here, we unveil an anisotropic ultrafast spin/valley dynamics in centimeter-scale, single-crystalline $T_d$-$WTe_2$ films using a femtosecond pump-probe technique at room temperature. We observe a transient (~0.8 ps) intra-valley transition and a subsequent polarization duration (~5 ps) during the whole spin/valley relaxation process. Furthermore, the relaxation exhibits the remarkable anisotropy of approximately six-fold and two-fold symmetries due to the intrinsic anisotropy along the crystalline orientation and the extrinsic matrix element effect, respectively. Our results offer a prospect for the ultrafast manipulation of spin/valleytronics in topological quantum materials for dissipationless high-speed spin/valleytronic devices.

*Introduction.*—Topological quantum materials have gained increasing interests owing to their remarkable new physics, which can be explored for their potential applications in the next generation novel devices and quantum computing [1-6]. They have experienced a tremendous expansion in popularity [7-9], which hosts anomalous linear response in the bulk and the nontrivial gapless surface states for dissipationless spintronics [10, 11]. In particular, the Weyl fermions of such surface states have exotic chiral behaviors, attracting worldwide attention in recent years [12-14]. $WTe_2$, a representative Weyl topological semimetal [15], which has type II Weyl points induced by the lack of inversion symmetry [16]. Its unique band structure and strong spin orbit coupling (SOC) give rise to a rich spectrum of exotic phases, including extremely large non-saturating magnetoresistance [17, 18], pressure-/gate-tunable superconductivity [19-22], the nonlinear Hall [23]/anomalous Hall effect [24], quantum spin Hall effect in monolayer [25, 26] and intriguing ferroelectric switching bi- or tri-layer structure [27]. Integration to the high-speed electronic devices, it becomes critical to study related ultrafast behaviors of $WTe_2$ within nano/picosecond time scale. The ultrafast carrier dynamics in $WTe_2$ has been investigated by time resolved reflectivity to manipulate charge degree of freedom [28, 29]. Moreover, Sie *et al.* reported an ultrafast switching of $WTe_2$ metastable phase based on lattice degree of freedom by field-induced lattice deformation through terahertz light pulses [30]. As such, great efforts will devote to the underlying ultrafast physics in $WTe_2$ for future potential applications.

Indeed, besides the aforementioned charge and lattice degrees of freedom, the compelling spin and valley textures offer new degrees of freedom for constructing devices for electronic/spintronic applications [31-33]. It often exists in monolayer transition metal dichalcogenides (TMDs) with direct bandgap located at the K/K′ points, which is attributed to the inversion symmetry breaking and strong SOC [34-37]. The opposite spin/valley index at K and K′ valleys with strong spin-valley interlocking strictly obey to the theory of polarization-dependent optical selection rules [38]. Thus, by a left/right polarized pulse laser, the carriers in the K/K′ valley can be selectively excited in these materials, which generate instant spin/valley polarization and subsequently experience an ultrafast relaxation process within nanosecond. Usually, time-resolved magneto-optical Kerr effect (TR-MOKE) based on the pump-probe technique is an effective approach to investigate these ultrafast spin/valley dynamics, and can continuously survey the signal of polarization on a long time scale beyond nanosecond. Since 2011, fascinating ultrafast spin relaxation in topological insulators (TIs) has been studied by pump-probe technique, i.e. $Bi_2Se_3$ [39, 40] and $Bi_2Te_2Se$ [41], indicating the prospect of topological quantum materials on this new degrees of freedom. Compared with TIs and monolayer TMDs, the additional broken inversion symmetry of $WTe_2$ enables further ultrafast exploration and manipulation of spin/valleytronics. Although the related spin/valley theoretical research by first principles calculations [42] and the ultrafast spin behavior in polycrystalline structure have been reported recently [29], the comprehensive research on the ultrafast spin/valley dynamics of single-crystal $WTe_2$ is still missing.

Here, we firstly reveal the ultrafast spin/valley dynamics of high-quality orthorhombic ($T_d$) phase of single-crystalline $WTe_2$ films at room temperature by TR-MOKE. The films were fabricated by pulsed laser deposition (PLD) technique as detailed in our previous reports [43, 44]. After the photoexcitation by circular polarization pump light, we demonstrate an intra-band transition with a lifetime of 0.8 ps when the spin/valley polarization direction of excited carriers is reversed. This transition corresponds to the spin split at the bottom of the valley of conduction band (CB). Subsequently, the polarization is exhausted by the further scattering under the assistance of inevitable defects with a lifetime of 5 ps, which is elucidated from the ultrafast transient reflectivity and the control experiments. More strikingly, the whole spin/valley relaxation process in $WTe_2$ exhibits a clear dual anisotropy, including an approximately six-fold and a two-fold symmetry by tuning the polarization orientation of the linear polarization probe light. The approximately six-fold symmetry is due to the intrinsic anisotropy of SOC along different orientations of $WTe_2$ crystal while the two-fold symmetry is related to the extrinsic matrix element effect. Our findings provide a unique insight into the potential manipulation of the spin/valley degree of freedom and applications of the room-temperature spin/valleytronic devices based on Weyl semimetal materials.

*The crystalline structure of high-quality $WTe_2$ films.*—The centimeter-scale single-crystalline $WTe_2$ films with thickness of 100 nm is fabricated by the modified PLD technique, which possesses a layered structure with an additional lattice distortion along the crystallographic axis *a* of tungsten chain [see Fig. 1(a)] [43, 44].

The typical Raman spectrum is shown in Supplemental Material [45], in which five peaks at about 110, 115, 132, 162 and 120 cm$^{-1}$ are observed, corresponding to $A_2^4$, $A_1^9$, $A_1^8$, $A_1^5$ and $A_1^2$ vibrational modes of $T_d$ phase $WTe_2$, respectively [51]. X-ray diffraction (XRD) spectrum in Supplemental Material [45] reveals the clear diffraction peaks corresponding to (002*n*) family of $WTe_2$, in good agreement with the previous results [52]. The selected-area electron diffraction (SAED) pattern further confirms the single-crystalline nature of $WTe_2$ films [Fig. 1(b)], where the corresponding lattice planes are marked. The atomic force microscope (AFM) surface morphology of $WTe_2$ films illustrates a roughness of 5.48 nm (see Supplemental Material [45]). The high-resolution transmission electron microscope (HRTEM) image further corroborates the high quality and single orientation by Fast Fourier Transform (FFT) analysis (see Supplemental Material [45]). In addition, the magnetotransport measurements show the clear Shubnikov-de Haas (SdH) quantum oscillations at low temperatures and the carrier concentration is calculated to be ~$10^{18}$ cm$^{-3}$ in our previous work[44].

*Ultrafast spin/valley dynamics revealed by the pump-probe technique.*—As mentioned before, $WTe_2$ has a layered structure with an additional lattice distortion along *a* axis (space group $P_{mn21}$). It makes an inversion symmetry breaking in single-crystal $WTe_2$. According to the requirements of time-reversal symmetry and lattice symmetry operations, the spin index along $-k_x$ and $k_x$ directions are opposite [29]. The illustration of the spin index along -X-Γ-X direction of momentum-space is shown in Fig. 1(c), where red and blue represent opposite polarization directions. Due to the inversion asymmetric property, a large spin-split is expected to exist at the bottom of CB. The theory of polarization-dependent optical selection rules plays an important role of our investigation, which indicates the left (right)-handed circularly polarized light $\sigma^+$ ($\sigma^-$) only photoexcite the carriers beneath the $-k_x$ ($k_x$) valley. After photoexcitation, the instant spin polarization of excited carriers can be monitored by time-resolved optical means, i.e. TR-MOKE. The schematic diagram of TR-MOKE set-up is shown in Fig. 2(a) (for details in Supplemental Material [45]). Here, we use 800 nm circularly polarized ($\sigma^+$ or $\sigma^-$) pulse to pump and 400 nm linearly polarized pulse to probe at room temperature. The Kerr rotation angle ($\theta_k$) of the polarization plane of the probe light is used to describe the detected spin polarization via the relationship of $\theta_k \sim \boldsymbol{M_p} \cdot \boldsymbol{k}$ (where $\boldsymbol{k}$ is the wave vector)[53]. We find that the Kerr signals change their signs by the different circularized ($\sigma^+$ and $\sigma^-$) pump light, as shown in Fig. 2(b). The difference of Kerr signals between $\sigma^+$ and $\sigma^-$ pump [$\Delta\theta_k(\sigma^+)$ - $\Delta\theta_k(\sigma^-)$] is defined as the net spin polarization signal, in order to rules out the artifacts from our measurements. In Fig. 2(c), we add an extrinsic in-plane magnetic field into this ultrafast process to check the variety of Kerr signals. Distinctly, with the magnetic field from 0 to 4000 Oe, the net Kerr signals show no obvious dependence or spin precession, indicating no Kerr Hanle effect existing. Like monolayer TMDs materials, the spin polarization seems to be “pinned” in the corresponding valley, which is due to the large out-of-plane effective magnetic field resulted from strong SOC effect [38]. Although a few works attribute the existence of Kerr Hanle effect to spins in localized defects [54, 55], the absence of Kerr Hanle effect undoubtedly demonstrates the potential coupling between spin and valley in $WTe_2$, almost the same as the previous report of polycrystalline structure [29]. In view of the semimetallic properties of $WTe_2$, the standard photoluminescence and related measurement of helicity are difficulty to be implemented. It indicates that the detected polarization signals cannot be verified totally attribute to valley polarization. Thus, we suppose that both spin and valley contribute to this Kerr signal, which is written as spin/valley polarization here.

In order to research the anisotropy of this spin/valley polarization, the polarization orientation of the probe light is changed. As shown in Fig. 2(a), $\varphi$ here represents the rotation angle with respect to the incident plane of probe light, which can be continuously changed in order to disclose the anisotropy of Kerr signals. In the case of $\varphi = 0^o$, the electric field of light lies in the plane of the incident probe beam, which is p-polarization (p-pol); for $\varphi = 90^o$, the probe light is s-polarization (s-pol). Through this procedure, the focus spot of pump beam is unchanged, assuring the Kerr signals more reliable. In Fig. 2(d), net Kerr signals $[\Delta\theta_k(\sigma^+) - \Delta\theta_k(\sigma^-)]$ versus time at the different $\varphi$ are displayed and the dashed lines represent the ground states. When $\varphi = 0^o$ (or $60^o$), the amplitude of net Kerr signal monotonically decreases and approaches the ground state with the lifetime of less than 1 ps. However, in the case of $\varphi = 10^o$ ($20^o$ or $90^o$), an obvious extreme point appears at the time-delay of ~6 ps: the amplitude of net Kerr signal initially reaches the minimum (maximum), then begins to increase (decrease) and finally gradually approaches the ground state. Therefore, the spin/valley polarization of excited carriers should experience two relaxation sub-processes, which is fitted by biexponential decay model (see

Supplemental Material [45]).

Fig. 2(e) shows the plot of $\varphi$ dependence of two relaxation sub-processes, where $A_1$ and $A_2$ denote the fitted amplitudes of each sub-process, respectively. Small error bars indicate the reliability of biexponential decay model. It can be found that the signs of $A_1$ and $A_2$ are always opposite, suggesting that a polarization reversal occurs. Considering the spin-split valley at the bottom of CB in $WTe_2$, this polarization reversal is attributed to an intra-band transition between two sub-bands with spin up or down. After the intra-band transition, a subsequent scattering sub-process by defects is demonstrated (see Supplemental Material [45]). The $\varphi$-dependent lifetime $\tau_1$ and $\tau_2$ are exhibited in Fig. 2(f), corresponding to intra-band transition and defects scattering sub-process, respectively. Almost unchanged $\tau_1$ and $\tau_2$ (average $\tau_1$ ~0.8 ps, $\tau_2$ ~5 ps) with respect to $\varphi$ verifies the isotropic relaxation rates ($\tau_1^{-1}$ and $\tau_2^{-1}$) of each sub-process.

The detailed $\varphi$ dependence of $A_1$ and $A_2$ is shown in Fig. 3(a), which is fitted from the original data in Supplemental Material [45]. Obviously regular symmetries of $A_1$ and $A_2$ can be observed. By double-sine function fitting (see Supplemental Material [45]), we can extract an approximately six-fold (~60°) and a two-fold (~180°) symmetry, clearly indicating a dual anisotropy of this ultrafast spin/valley relaxation in $WTe_2$. To reveal the origin of this dual anisotropy, we investigate the dependent net Kerr signals by rotating the a-axis of $WTe_2$ film as a control experiment. Here $\alpha$ is defined as the angle between the crystallographic axis [100] of the $WTe_2$ film and the spatial $x$ axis [as marked in Fig. 2(a)]. The fitted $A_1$ ($A_2$) as a function of $\alpha$ only exhibits an approximately six-fold symmetry (~60°) [in Fig. 3(b)], exactly matching one of the symmetries (~60°) obtained from Fig. 3(a). The polar coordinate of $A_2$ as a function of $\alpha$ is illustrated in Fig. 3(c) for more intuitive display of this six-fold symmetry. It indicates that this approximately six-fold symmetry is related to the anisotropy of $WTe_2$ crystal. Furthermore, we also noticed that the anisotropic SOC along different crystallographic orientations can influence the behaviors of spin/valley polarization [56]. Thus, we suppose that this approximately six-fold symmetry of spin/valley polarization originates from the anisotropic SOC in $WTe_2$. In addition, $WTe_2$ does not have a standard six-fold symmetry due to the lattice distortion in tungsten chain [Fig. 1(a)], which generates the non-standard component of six-fold symmetry in Fig. 3(b).

Next, we subtract the approximately six-fold symmetry from Fig. 3(a) and plot the rest two-fold symmetry of spin/valley polarization in Supplemental Material [45]. Since this two-fold symmetry does not depend on $\alpha$, it may arise from the measurement approach itself by tuning polarization orientation of the probe light rather than the intrinsic anisotropy of this ultrafast spin/valley relaxation. Considering the matrix element effects, s-pol or p-pol probe light may interact with the different areas of energy bands [57]. It is extracted from the matrix element of transitions between band states at the extrema, determining the allowed and forbidden transitions, which indicates that the different polarization of the light can access to the different positions in the momentum space. This allows that the s-pol or p-pol probe light detects different band structure textures along the same momentum space direction, also revealed by angle resolved photoelectron spectroscope (ARPES) [58]. For ultrafast spin/valley dynamics in $WTe_2$, the s-pol or p-pol probe light may actually detect two different areas of the momentum space. As a result, the two-fold symmetry of spin/valley polarization is due to the varied s-pol or p-pol component of the probe light upon its polarization orientation ($\varphi$).

*Additional control experiments and discussion.*—Finally, the fluence dependences of the Kerr signals and ultrafast transient reflectivity are also researched, as shown in Supplemental Material [45]. It can be revealed that the ultrafast spin/valley dynamics of $WTe_2$ is a single-photon process and a large excited electronic density of state exists in $WTe_2$ films. Moreover, in order to exclude other potential effects, more additional results of ultrafast transient reflectivity and Kerr signals of $WTe_2$ are shown in Supplemental Material [45].

Generally, a model based on band structure is proposed to describe this ultrafast spin/valley dynamics. The band structure model is schematically shown in Fig. 4. Due to the inversion asymmetric property in $WTe_2$, the valley at bottom of CB is split into two sub-bands with the opposite polarization [29]. The carriers in valence band (VB) are photogenerated by $\sigma^+$ pump pulse and then transfer to CB with spin/valley polarization up obeying the polarization-dependent optical selection rules. Firstly, the excited carriers instantly decay to Position 'A' via the intra-band-particle interaction. Such a process usually lasts for about tens of femtoseconds escaping from our detection limit. Secondly, carriers with up-polarization transfer to the lower sub-band ($\Delta E \sim 10$ meV) [29] at a rate $\tau_1^{-1}$ with down-polarization at Position 'B', which is also an intra-band transition. This sub-process is governed by Elliot-Yafet mechanism with the participation of electrons [47], similar to the spin-flip transition in $WS_2$ [59]. Eventually, the down-polarization is exhausted at a rate $\tau_2^{-1}$ due to the defect-scattering sub-process.

*Conclusion.*—In summary, we have revealed the room-temperature ultrafast spin/valley dynamics by TR-MOKE technique in centimeter-scale, single-crystalline $T_d$-$WTe_2$ films with low carrier concentration of ~$10^{18}$ cm$^{-3}$. A band structure model is proposed to illustrate the whole ultrafast spin/valley relaxation, including intra-band polarization-flip transition and defect-scattering sub-process at the order of picosecond. The intra-band polarization-flip transition corresponds to the spin split at the bottom of the valley of CB, which has potential applications for the multichannel high-speed quantum computing. More importantly, we have unveiled the dual anisotropy, including an approximately six-fold and a two-fold symmetry, of ultrafast spin/valley relaxation. The former is an intrinsic symmetry arising from the anisotropy of SOC along different orientations of $WTe_2$ crystal. The latter is due to the probe condition related to the matrix element effect. Our findings not only enrich the knowledge of the spin/valleytronics in topological quantum materials, but also advance a novel route towards future spin/valleytronics Weyl devices.

We would like to acknowledge the stimulating discussions with Prof. Jingbo Qi and Prof. Xiangang Wan. This work was partially supported by the National Key R&D Program of China (Grant Nos. 2017YFA0206304 and 2016YFA0300803), the

National Natural Science Foundation of China (Grant Nos. 61822403, 11874203, 11774160, 61427812, and U1732159), the Fundamental Research Funds for the Central Universities (Grant. Nos. 021014380080), the Natural Science Foundation of Jiangsu Province of China (Grant. No. BK20192006) and Collaborative Innovation Center of Solid-State Lighting and Energy-Saving Electronics.

Y. C. and Z. C. contributed equally to this work. X. W. and Y. X. conceived the idea and supervised the project.

*xfwang@nju.edu.cn

#ybxu@nju.edu.cn

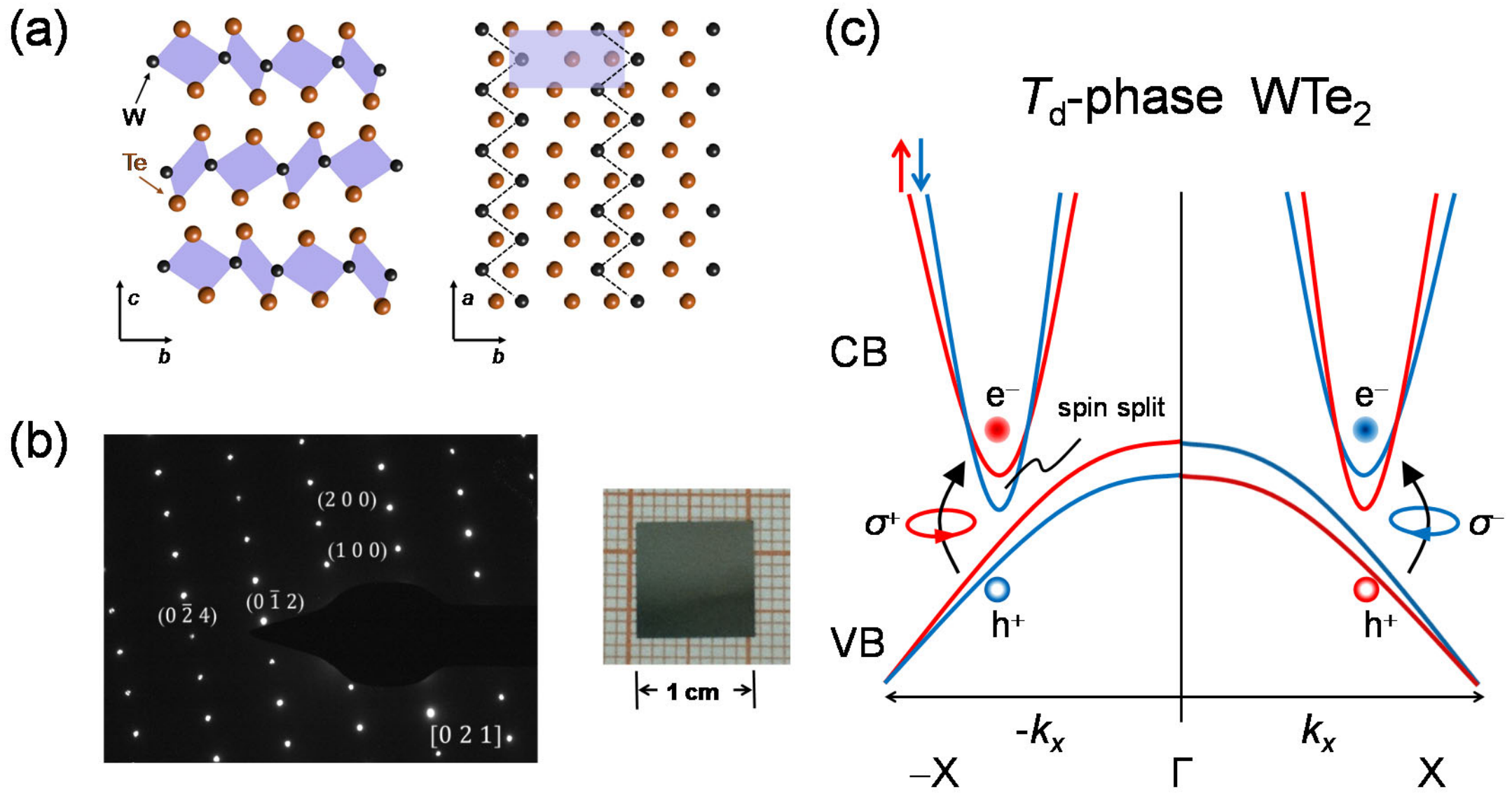

FIG. 1. Structural characterization of $T_d$-$WTe_2$ films (a) Lattice structure: side view (b-c plane) and top view (a-b plane). The shaded area exhibits the unit cell. The dashed lines indicate the W-W zigzag chain along *a* axis. (b) SAED pattern taken along the [021] zone axis (left) and a photograph of the centimeter-scale $WTe_2$ film on the mica substrate (right). (c) Diagrammatic sketch of spin index along -X-Γ-X direction of momentum-space in $WTe_2$, where red and blue represent spin up and down, respectively.

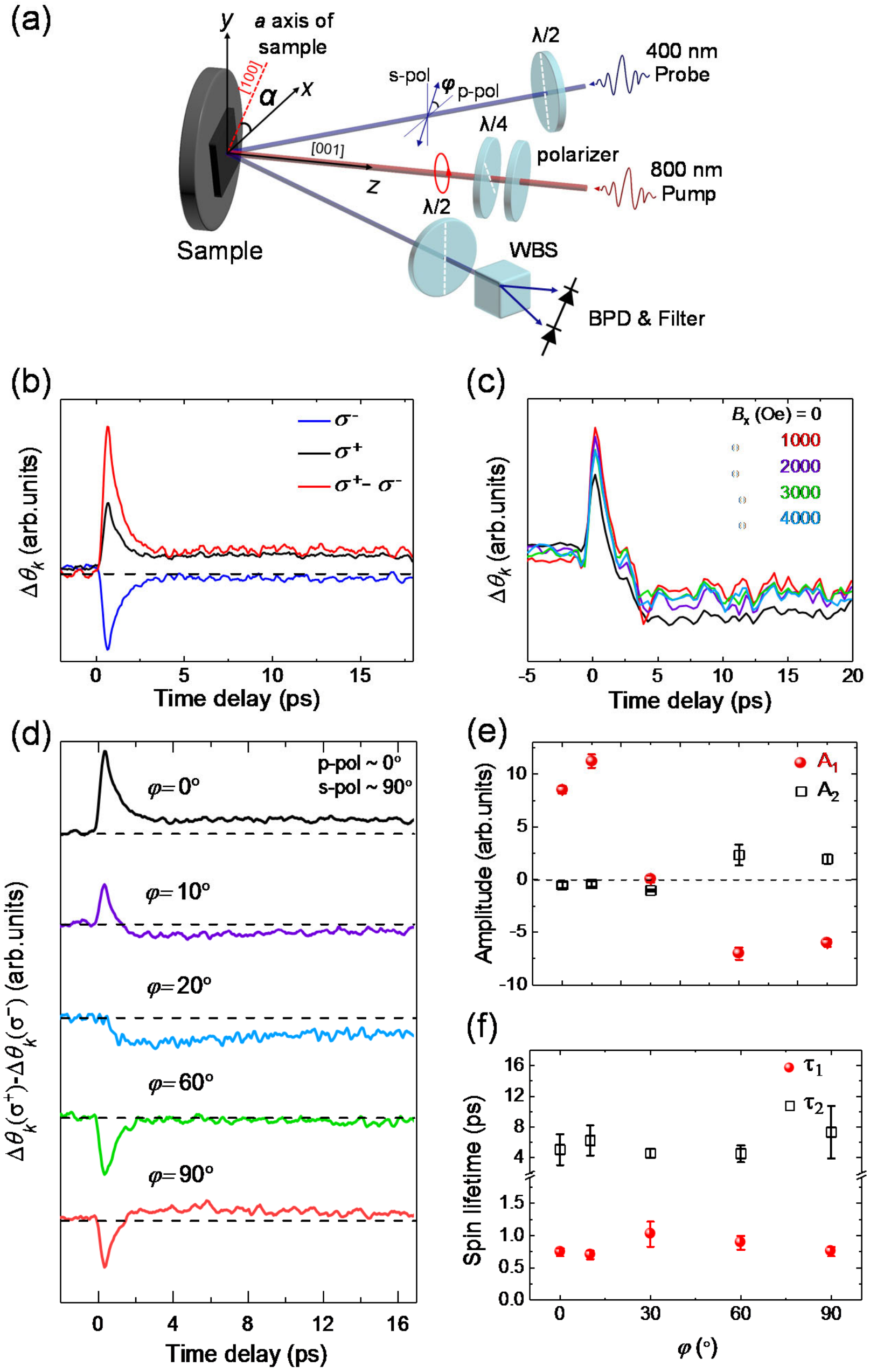

FIG. 2. Ultrafast spin/valley dynamics in $WTe_2$ films (a) Schematic diagram of TR-MOKE setup. $\varphi$ represents the rotation angle with respect to the incident plane of probe light. In the case of $\varphi = 0^o$, the electric field of light lies in the plane of the incident probe beam, which is p-pol. $\alpha$ is defined as the angle between the crystallographic axis [100] of the $WTe_2$ film and the spatial $x$ axis. $\lambda/2$: half-wave plate. $\lambda/4$: quarter-wave plate. (b) Time-resolved Kerr rotation traces under excitation of $\sigma^+$ and $\sigma^-$ pump along with the signals difference [$\Delta\theta_k(\sigma^+)$ - $\Delta\theta_k(\sigma^-)$] between $\sigma^+$ and $\sigma^-$ pump. (c) The net Kerr signals as a function of delay time at different external magnetic field (from 0 ~ 4000 Oe). (d) The net Kerr signals as a function of delay time at different $\varphi$. (e,f) The $\varphi$ dependent fitted $A_1$, $A_2$ and spin lifetimes ($\tau_1$ and $\tau_2$) from the net Kerr signals. The average values of $\tau_1$ and $\tau_2$ are 0.8 ps and 5 ps, respectively.

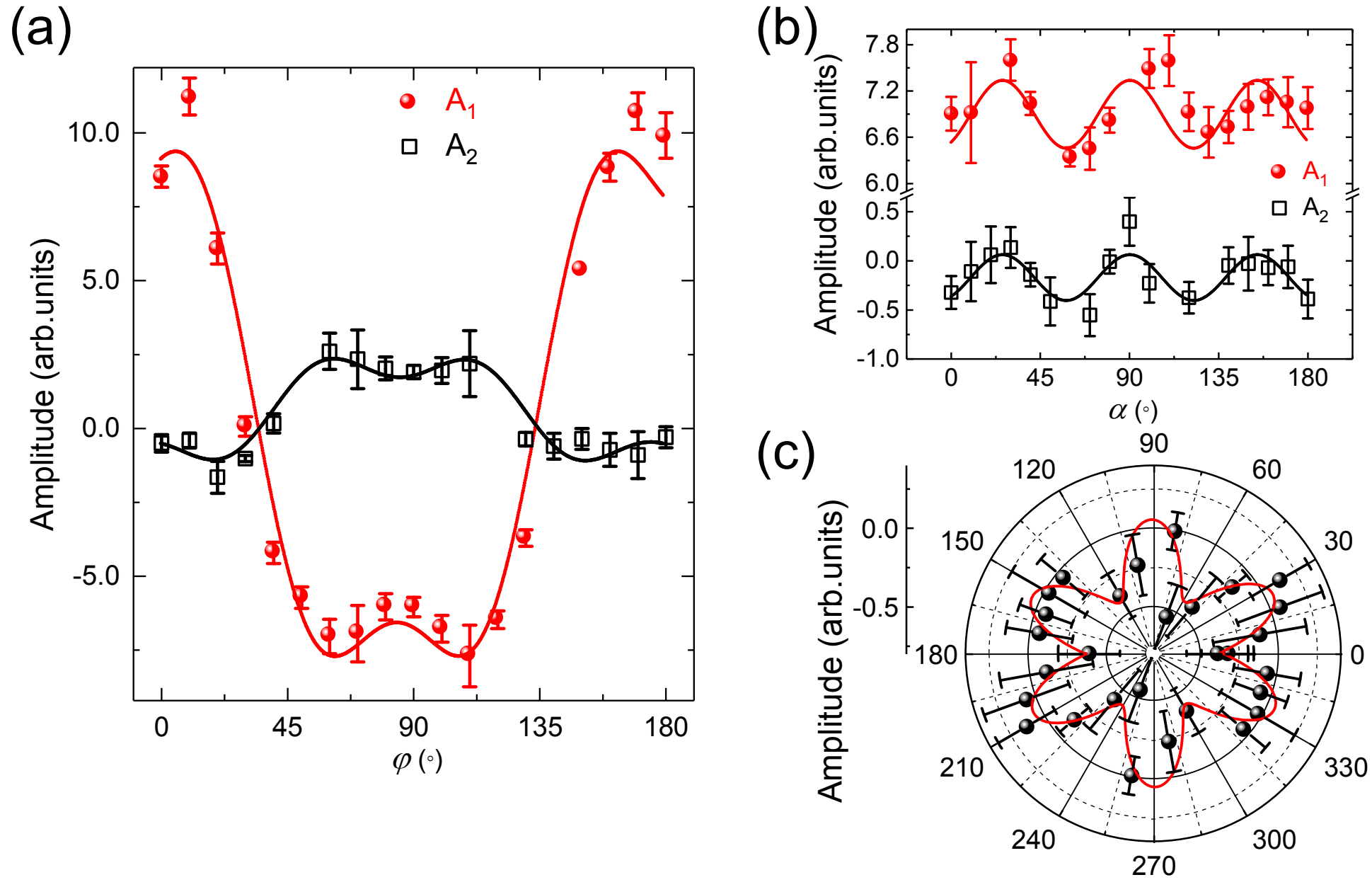

FIG. 3. Anisotropy of ultrafast spin/valley dynamics (a) The $\varphi$ dependence of fitted $A_1$, $A_2$ from the net Kerr signals in Supplemental Material [45]. Here $\varphi$ is ranged from 0° to 180° while Fig. 2(d) only shows a small portion of the data from 0° to 90°. A few points are ruled out when the values of $A_1$ or $A_2$ are close to zero for their very large error bars. The red and black curves are the fitted lines by double-sine function. (b) The $\alpha$ dependence of fitted $A_1$ and $A_2$. $\alpha$ is also ranged from 0° to 180°. (c) The polar coordinate of $A_2$ as a function of $\alpha$. The points from 180° to 360° are the symmetric manipulation of the data from 0° to 180°. Two points at 90° and 270° are ignored in order to present the approximately six-fold symmetry more clearly. The red and black curves in (b) and (c) are the fitted lines by sine function.

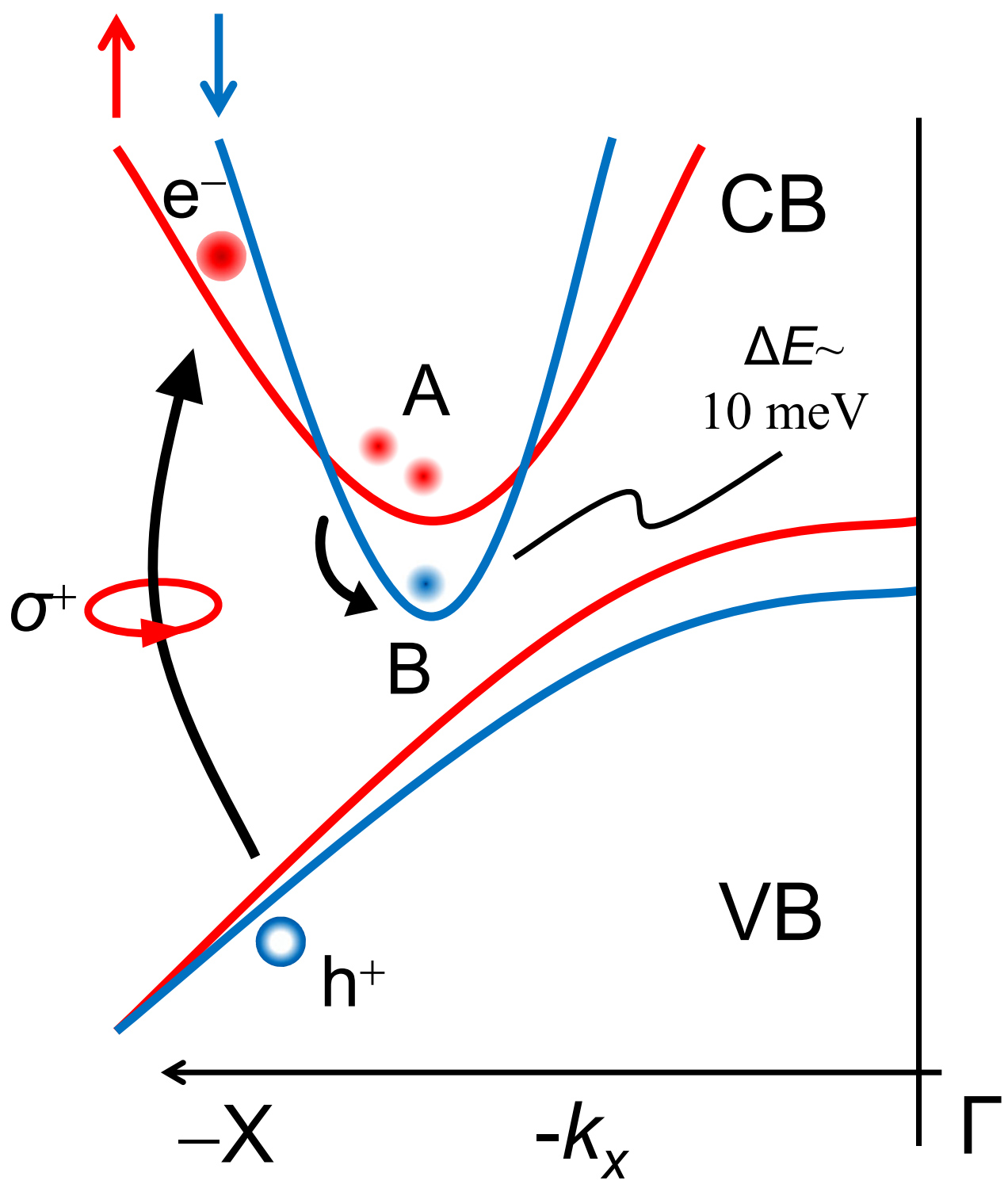

FIG. 4. Schematic band structure of $WTe_2$ with ultrafast spin/valley dynamics: blue and red denote different polarized states, respectively. The hollow and solid circles represent holes and electrons, respectively. Symbol ‘A’ denotes the position close to the bottom of the conduction sub-band with up-polarization. Symbol ‘B’ denotes the position close to the bottom of the conduction sub-band with down-polarization.